\documentclass[aps,prl,twocolumn,superscriptaddress,showpacs,floatfix]{revtex4}
\usepackage{graphicx}
\usepackage{epsfig}
\usepackage{amsmath}
\usepackage{amssymb}
\usepackage{natbib}
\usepackage{color}

\begin{document}

\title{Nature of Partial Magnetic Order in the Frustrated Antiferromagnet Gd$_{2}$Ti$_{2}$O$_{7}$}

\author{Joseph A. M. Paddison}
\affiliation{Department of Chemistry, University of Oxford, Inorganic Chemistry Laboratory, South Parks Road, Oxford OX1 3QR, U.K.}
\affiliation{ISIS Facility, Rutherford Appleton Laboratory, Harwell Oxford, Didcot OX11 0QX, U.K.}

\author{Andrew B. Cairns}
\affiliation{Department of Chemistry, University of Oxford, Inorganic Chemistry Laboratory, South Parks Road, Oxford OX1 3QR, U.K.}

\author{Dmitry D. Khalyavin}
\affiliation{ISIS Facility, Rutherford Appleton Laboratory, Harwell Oxford, Didcot OX11 0QX, U.K.}

\author{Pascal Manuel}
\affiliation{ISIS Facility, Rutherford Appleton Laboratory, Harwell Oxford, Didcot OX11 0QX, U.K.}

\author{Aziz Daoud-Aladine}
\affiliation{ISIS Facility, Rutherford Appleton Laboratory, Harwell Oxford, Didcot OX11 0QX, U.K.}

\author{Georg Ehlers}
\affiliation{Quantum Condensed Matter Division, Oak Ridge National Laboratory, Oak Ridge, Tennessee 37831, USA}

\author{Oleg A. Petrenko}
\affiliation{Department of Physics, University of Warwick, Coventry CV4 7AL, U.K.}

\author{Jason S. Gardner}
\affiliation{Neutron Group, National Synchrotron Radiation Research Center, Hsinchu, Taiwan 30076}

\author{H. D. Zhou}
\affiliation{Department of Physics and Astronomy, University of Tennessee, Knoxville, Tennessee 37996-1200, USA}
\affiliation{National High Magnetic Field Laboratory, Florida State University, Tallahassee, Florida 32310-3706, USA}

\author{Andrew L. Goodwin}
\affiliation{Department of Chemistry, University of Oxford, Inorganic Chemistry Laboratory, South Parks Road, Oxford OX1 3QR, U.K.}

\author{J. Ross Stewart}
\affiliation{ISIS Facility, Rutherford Appleton Laboratory, Harwell Oxford, Didcot OX11 0QX, U.K.}

\date{\today}

\begin{abstract}
The frustrated pyrochlore antiferromagnet Gd$_{2}$Ti$_{2}$O$_{7}$ has an unusual partially-ordered magnetic structure at the lowest measurable temperatures. This structure is currently believed to involve four magnetic propagation vectors $\mathbf{k}\in \langle \frac{1}{2} \frac{1}{2} \frac{1}{2} \rangle^*$ in a cubic 4-$\mathbf{k}$ structure, based on analysis of magnetic diffuse-scattering data [\emph{J. Phys.: Condens. Matter} \textbf{16}, L321 (2004)]. Here, we present three pieces of evidence against the 4-$\mathbf{k}$ structure. First, we report single-crystal neutron-diffraction measurements as a function of applied magnetic field, which are consistent with the selective field-induced population of non-cubic magnetic domains. Second, we present evidence from high-resolution powder neutron-diffraction measurements that rhombohedral strains exist within magnetic domains, which may be generated by magneto-elastic coupling only for the alternative 1-$\mathbf{k}$ structure. Finally, we show that the argument previously used to rule out the 1-$\mathbf{k}$ structure is flawed, and demonstrate that magnetic diffuse-scattering data can actually be fitted quantitatively by a 1-$\mathbf{k}$ structure in which spin fluctuations on ordered and disordered magnetic sites are strongly coupled. Our results provide an experimental foundation on which to base theoretical descriptions of partially-ordered states.

\end{abstract}

\pacs{75.10.Jm,61.43.Bn,75.25.-j}
\maketitle

Geometrical frustration is a central theme of condensed-matter physics because it generates exotic magnetic states. These states can usually be divided into two categories: spin liquids, in which frustration inhibits long-range magnetic order, and spin solids, in which perturbations to the dominant frustrated interactions eventually generate magnetic order \cite{Moessner_2006}. Defying this classification, some frustrated magnets exhibit \emph{partial} magnetic order \cite{Movshovich_1999,Greedan_2002,Zheng_2005,Rule_2007,Cao_2007,Ehlers_2007,Hayes_2011}, in which only some spins order at low temperature. Partial magnetic order is surprisingly prevalent in frustrated systems \cite{Movshovich_1999,Greedan_2002,Zheng_2005,Rule_2007,Cao_2007,Ehlers_2007,Hayes_2011} and partial structural order plays a key role in determining the properties of materials such as fast-ion conductors and high-pressure elemental phases \cite{Keen_1996,Gregoryanz_2008}. The motivation for studying partial order lies in understanding the interplay between order and disorder that may drive its formation \cite{Chern_2008,Javanparast_2015,Wills_2006}. To achieve this aim, experimental determination of the nature of partial order in a real material is essential.

\begin{figure}
\begin{center}
\includegraphics[scale=1]{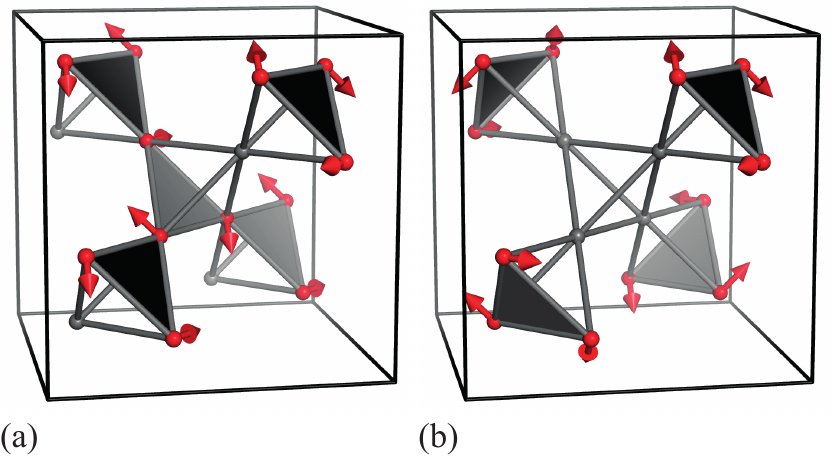}
\end{center}
\caption{\label{structures} (a) 1-$\mathbf{k}$ and (b) 4-$\mathbf{k}$ magnetic structure candidates for Gd$_{2}$Ti$_{2}$O$_{7}$. In both structures, the ordered site is shown in red, the disordered site is shown in grey, and average spin orientations are shown as red arrows. A single crystallographic unit cell is shown; spin orientations are reversed in adjacent unit cells.}\end{figure}

The frustrated antiferromagnet Gd$_{2}$Ti$_{2}$O$_{7}$ is a canonical partially-ordered system in which magnetic Gd$^{3+}$ ions ($S=7/2$) occupy a pyrochlore network of corner-sharing tetrahedra \cite{Gardner_2010}. Two magnetic phase transitions occur at $T_{1} = 1.1$\,K and $T_{2} = 0.75$\,K \cite{Raju_1999,Ramirez_2002,Bonville_2003,Petrenko_2004,Petrenko_2011} to different states with magnetic propagation vector $\mathbf{k}=[\frac{1}{2} \frac{1}{2} \frac{1}{2}]^*$ \cite{Champion_2001,Stewart_2004}. 
Here, we consider only the low-temperature ($T \ll T_2$) state. Two different magnetic structures are equally consistent with most experimental data, including local probes \cite{Bonville_2003,Sosin_2008,Sosin_2006} and the magnetic Bragg scattering observed in powder neutron-diffraction experiments \cite{Champion_2001,Stewart_2004}. These candidate magnetic structures are shown in Fig.~\ref{structures}. The ``1-$\mathbf{k}$ structure'' involves a single $\mathbf{k}=[\frac{1}{2} \frac{1}{2} \frac{1}{2}]^*$ while the ``4-$\mathbf{k}$ structure" involves a superposition of the four symmetry-equivalent $\mathbf{k}\in \langle \frac{1}{2} \frac{1}{2} \frac{1}{2} \rangle^*$. In both structures, 3/4 of the spins participate in long-range order, while the remaining spins show mostly short-range (paramagnetic) correlations with only a small ordered magnetic moment \cite{Stewart_2004}. Yet, the nature of partial order is very different for the two structures: the 1-$\mathbf{k}$ structure has rhombohedral symmetry (magnetic space group $R_{\mathrm{I}}\bar{3}m$) and disordered spins are separated by 7.2\,\AA, while the 4-$\mathbf{k}$ structure has cubic symmetry (magnetic space group $F_{\mathrm{S}}\bar{4}3m$) and disordered spins are separated by 3.6\,\AA. 
Evidence for the magnetic structure of Gd$_{2}$Ti$_{2}$O$_{7}$ has so far been indirect: it was argued in Ref.~\onlinecite{Stewart_2004} that the magnetic diffuse scattering observed in powder neutron-diffraction experiments is consistent only with the smaller distance between disordered spins in the 4-$\mathbf{k}$ structure.

In this Letter, we present experimental evidence that the low-temperature magnetic structure of Gd$_{2}$Ti$_{2}$O$_{7}$ is 1-$\mathbf{k}$, challenging the 4-$\mathbf{k}$ structure that was previously proposed \cite{Stewart_2004}. We employ a combination of single-crystal and powder neutron diffraction to demonstrate the existence of magnetic domains with non-cubic symmetry, consistent with the 1-$\mathbf{k}$ structure. We further show that ordered and disordered sites are not independent as was previously assumed \cite{Stewart_2004}, but are actually coupled via spin fluctuations.

Direct evidence against the 4-$\mathbf{k}$ structure is obtained from single-crystal neutron-diffraction measurements. Magnetic domains of the 1-$\mathbf{k}$ structure select different $\mathbf{k}\in \langle \frac{1}{2} \frac{1}{2} \frac{1}{2} \rangle^*$ and therefore have different single-crystal diffraction patterns, while 4-$\mathbf{k}$ domains are related only by translational and time-reversal symmetries and hence appear identical to neutrons. However, a macroscopic 1-$\mathbf{k}$ sample is expected to contain equal populations of the four degenerate $\mathbf{k}\in \langle \frac{1}{2} \frac{1}{2} \frac{1}{2} \rangle^*$ domains, which would yield identical magnetic Bragg scattering to the 4-$\mathbf{k}$ structure. We addressed this problem by applying a weak magnetic field to our sample in order to break the domain degeneracy. 
The WISH diffractometer \cite{Chapon_2011} at the ISIS neutron source was used to measure magnetic Bragg peaks at $T=0.07$\,K with a magnetic field applied along the $[1\bar{1}0]$ direction. The sample was a single crystal of volume $\sim$10\,$\mathrm{mm^{3}}$, which was cut from a larger crystal of 99.4\% isotopically-enriched $^{160}$Gd$_{2}$Ti$_{2}$O$_{7}$ prepared by the floating-zone image furnace method \cite{Balakrishnan_1998,Gardner_1998} using an isotopically-natural Gd$_{2}$Ti$_{2}$O$_{7}$ seed crystal.

\begin{figure}
\begin{center}
\includegraphics[scale=1]{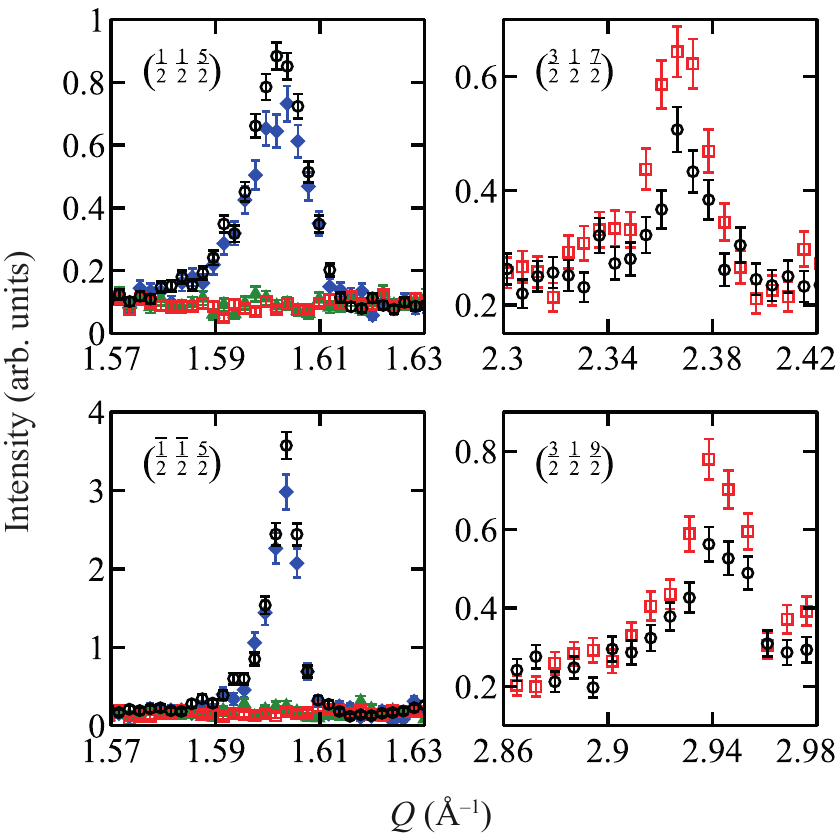}
\end{center}
\caption{\label{wish_data} Intensities of selected single-crystal magnetic Bragg peaks at different values of the applied magnetic field $\mu_{0}\mathbf{H}$ parallel to $[1\bar{1}0]$ at $T = 0.07$\,K. Magnetic Bragg peaks are labelled in each panel; left-hand panels show peaks within the $(hhl)$ plane and right-hand panels show peaks outside the $(hhl)$ plane. Points are coloured as follows: $\mu_{0}\mathbf{H}=0$ (black hollow circles), $\mu_{0}\mathbf{H}=0.1$\,T (blue filled diamonds), $\mu_{0}\mathbf{H}=0.2$\,T (green filled triangles), and $\mu_{0}\mathbf{H}=0.5$\,T (red hollow squares). For clarity, only $\mu_{0}\mathbf{H}=0$ and $\mu_{0}\mathbf{H}=0.5$\,T are shown in the right-hand panels.}\end{figure}

The field-dependence of selected magnetic Bragg peaks is shown in Fig.~\ref{wish_data}. 
Magnetic Bragg peaks in the $(hhl)$ plane disappear on the application of a
magnetic field, while the intensity of magnetic Bragg peaks outside the $(hhl)$ plane increases. This result is
consistent with the 1-\textbf{k} structure, because the applied field
makes the same angle ($90^{\circ}$) with two of the $\mathbf{k}$ which lie within the $(hhl)$ plane, and a different angle ($35^{\circ}$) with the remaining two $\mathbf{k}$ which lie outside the $(hhl)$ plane (see SI). Hence one pair of domains is populated by the application of a
field and the other pair is depopulated.  
The only scenario in which our data might be consistent with the 4-$\mathbf{k}$ structure
is if the applied magnetic field actually caused a magnetic phase
transition rather than a domain imbalance. This is unlikely, because there is no
experimental evidence for a field-induced phase transition in either specific heat \cite{Petrenko_2004} or torque magnetometry \cite{Glazkov_2007} measurements
for fields along $\left\langle 110\right\rangle $ of less than $2\,\mathrm{T}$, whereas we observe changes in peak intensity for much smaller fields
$B\approx0.2$\,T. However, the evidence for field-induced spin reorientations in Er$_2$Ti$_2$O$_7$ \cite{Ruff_2008} motivated further experimental work, which we discuss below.

We now show that a signature of non-cubic domain structure persists in zero applied field. If the magnetic structure is 1-$\mathbf{k}$, magneto-elastic coupling may generate a rhombohedral lattice distortion (see, e.g., \cite{Goodwin_2006}), whereas no such distortion is possible for the 4-$\mathbf{k}$ structure.
To look for a distortion, we measured the powder neutron-diffraction pattern of Gd$_{2}$Ti$_{2}$O$_{7}$ at $T=1.1$\,K and $T=0.03$\,K. The HRPD diffractometer at ISIS was used because it has among the highest-available reciprocal-space resolution ($\Delta Q/Q \approx 5\times 10^{-5}$). The powder sample was the same as used in previous studies \cite{Champion_2001,Stewart_2004} and was mounted in a Cu holder to which deuterated isopropyl alcohol (d-IPA) was added to improve thermal contact. A dilution refrigerator was used to reach sub-Kelvin temperatures. We performed a series of Rietveld refinements to the HPRD data in which the extent of rhombohedral distortion was systematically varied. This was achieved by defining a distortion parameter, $D=c_{\mathrm{h}}/\sqrt{6}a_{\mathrm{h}} -1$, where $a_{\mathrm{h}}$ and $c_{\mathrm{h}}$ are lattice parameters of the crystallographic unit cell in the hexagonal setting of space group $R\bar{3}m$. For the undistorted cubic structure, $D=0$. Rietveld refinements were performed using \textsc{Topas} Academic (version 5) \cite{Coelho_2012} and the \textsc{Isodistort} software was used to generate structural models \cite{Campbell_2006,Stokes_2006}. The Bragg peaks arising from Cu and V were fitted by Pawley refinement and the unit-cell volume, sample peak-shape, and background parameters were refined. 
 
\begin{figure}
\begin{center}
\includegraphics[scale=1]{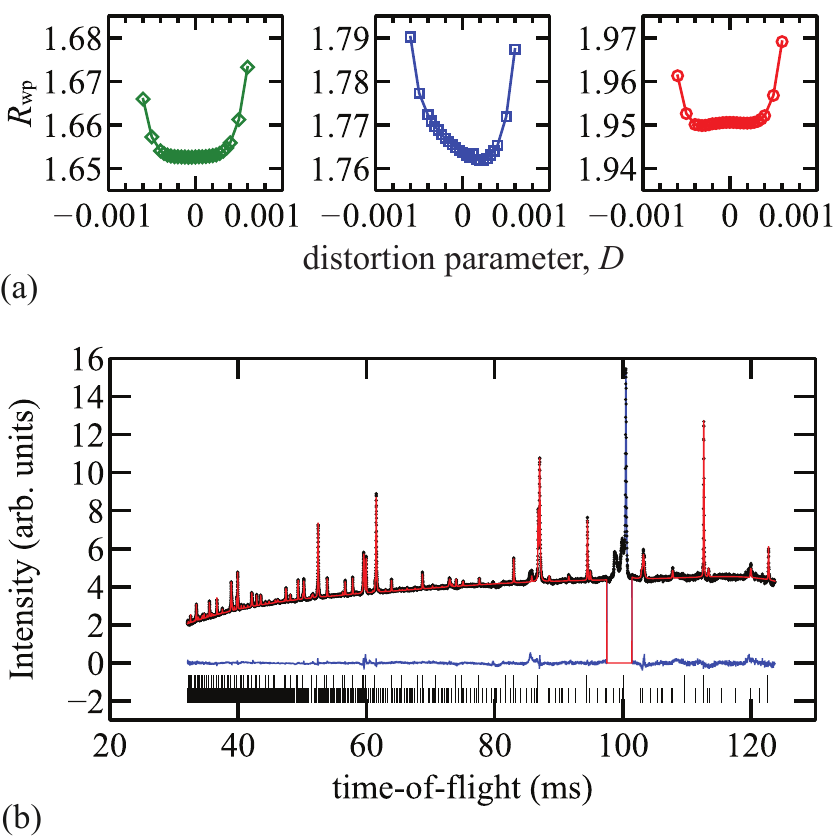}
\end{center}
\caption{\label{hrpd} (a) Dependence of goodness-of-fit $R_\mathrm{wp}$ on the distortion parameter $D$ (defined in the text) for Rietveld refinement to high-resolution powder neutron-diffraction data. Green diamonds show refinement of the nuclear phase to $T=1.1$\,K\,($>T_{1}$) data, blue squares refinement of the magnetic+nuclear phase to $T=0.03$\,K\,($\ll T_{2}$) data, and red circles refinement of the nuclear phase only to $T=0.03$\,K data. (b) High-resolution powder neutron-diffraction patterns at $T=0.03$\,K. Experimental data are shown as black circles, Rietveld fit as a red line, and data--fit as a blue line. The upper line of tick marks indicates the positions of nuclear Bragg peaks and the lower line of tick marks indicates the positions of magnetic Bragg peaks. The region around 100\,ms was excluded as it contains a large background contribution from d-IPA that could not be robustly refined.}\end{figure}

The dependence of the goodness-of-fit (described by the weighted-profile $R$-factor $R_{\mathrm{wp}}$) on $D$ is shown in Fig.~\ref{hrpd}(a). For refinement to the $T=1.1$\,K data, $R_{\mathrm{wp}}$ has a symmetric minimum around $D=0$; hence no lattice distortion exists in the paramagnetic phase within the resolution of our data. For refinement to the $T=0.03$\,K data, we fixed the magnetic structure to be 1-$\mathbf{k}$ (we note that this does not bias the results, since the 1-$\mathbf{k}$ and 4-$\mathbf{k}$ structures have identical powder-averaged Bragg scattering for $D=0$ \cite{Champion_2001}). We allowed two additional parameters to refine to fit the magnetic phase: the size of the ordered magnetic moment $\mu_{\mathrm{ord}}$ for the ordered 3/4 of the spins, and a Gaussian broadening parameter for the magnetic peaks that corresponds to a finite magnetic-domain size $\xi_\mathrm{domain}$. Our key result is that the minimum $R_{\mathrm{wp}}$ is now obtained for a non-zero rhombohedral distortion $D=0.00022(4)$ with lattice parameters $a_{\mathrm{h}} = 7.19352(9)$\,\AA~and $c_{\mathrm{h}} = 17.6244(6)$\,\AA. The fit-to-data for this model is shown in Fig.~\ref{hrpd}(b). The rhombohedral distortion is statistically-significant: $|c_{\mathrm{h}}-c^{0}_{\mathrm{h}}|/\sigma(c_{\mathrm{h}})=4$, where $c^{0}_{\mathrm{h}}=17.62177(3)$\,\AA~is obtained for refinement to the $T=0.03$\,K data with the cubic constraint that $D=0$. We find $\mu_\mathrm{ord}=6.7(1)\,\mu_\mathrm{B}$ for the ordered site, consistent with previous studies \cite{Champion_2001,Stewart_2004}. Finally, we investigate whether the sensitivity to this distortion arises primarily from the magnetic or nuclear peaks, by performing a further set of refinements to the $T=0.03$\,K data where the magnetic phase was not included in the model. We find that the sensitivity of the data to a rhombohedral distortion is greatly reduced for these refinements. This result suggests that the rhombohedral distortion does not persist over the length-scale of the powder crystallites but only over the finite magnetic-domain size $\xi_\mathrm{domain} = 2.4(2)\times 10^{3}$\,\AA---an unusual behaviour that has also been reported for the small monoclinic distortion present in MnO at low temperature \cite{Goodwin_2006,Frandsen_2015}.

Our observations of rhombohedral symmetry-breaking apparently contradict a previous analysis \cite{Stewart_2004}, where the distances between disordered spins in the 1-$\mathbf{k}$ structure were shown to be incompatible with the presence of a broad magnetic diffuse-scattering peak at $Q\approx 1.1$\,\AA$^{-1}$. Implicit in this argument is the assumption that the ordered site does not give rise to diffuse scattering. While this assumption may appear plausible, it is actually flawed, as we now show. The integral of the magnetic scattering intensity over all $Q$ and energy, $I_\mathrm{int}$, is proportional to the total squared magnetic moment, $\mu^{2}=(g\mu_{\mathrm{B}})^{2}S(S+1)$, which is temperature-independent \cite{Jensen_1991}. For a site which shows long-range spin order, one component of spin is aligned along a local axis with maximum projection $\pm S$ at low temperature, generating Bragg scattering with intensity proportional to $\mu_{\mathrm{ord}}^{2}=(g\mu_{\mathrm{B}})^{2}S^{2}$. The transverse spin components undergo fluctuations (e.g., spin waves) and contribute the rest of $I_\mathrm{int}$ \cite{Aeppli_1999,Squires_1978}. In a diffraction experiment, an energy integral over spin fluctuations is performed, so that magnetic diffuse scattering is present even in conventionally-ordered magnets at low temperature \cite{Mellergard_1998,Mellergard_1999}. In Gd$_{2}$Ti$_{2}$O$_{7}$, we estimate the proportion of the diffuse scattering which arises from ordered and disordered sites by taking $\mu=7.94\,\mu{_\mathrm{B}}$ for both sites and assuming that the ordered site has its maximum $\mu_{\mathrm{ord}}=7.0\,\mu{_\mathrm{B}}$ whereas the disordered site has $\mu_{\mathrm{ord}}=0$. Then, 25\% of $I_\mathrm{int}$ will be diffuse scattering from the disordered site, and $3/[4(S+1)] = 17\%$ of $I_\mathrm{int}$ will be diffuse scattering from the ordered site. Hence, the net contribution to the diffuse scattering from ordered and disordered sites is actually expected to be similar.

We now develop a quantitative model of the partial-order state which is consistent with all our previous results.  We use reverse Monte Carlo (RMC) refinement \cite{Mellergard_1998,Mellergard_1999,Paddison_2012,Paddison_2013} to determine whether spin configurations exist that obey three constraints. (i) A single model is consistent with the total (Bragg+diffuse) magnetic neutron-scattering data.  (ii) Each Gd$^{3+}$ ion has the same $\mu$. (iii) The component of the magnetic structure that shows long-range order (``average structure") is 1-$\mathbf{k}$. Spin configurations were refined to the powder-diffraction data of Ref.~\onlinecite{Stewart_2004}, which were collected at $T=0.25$\,K using the D20 diffractometer at the Institut Laue-Langevin \cite{Hansen_2008}. Nuclear Bragg peaks and non-magnetic background were removed from the data before refinement; full details of the data processing and RMC refinement algorithm are given in SI.

\begin{figure}[htpb!]
\begin{center}
\includegraphics[scale=1]{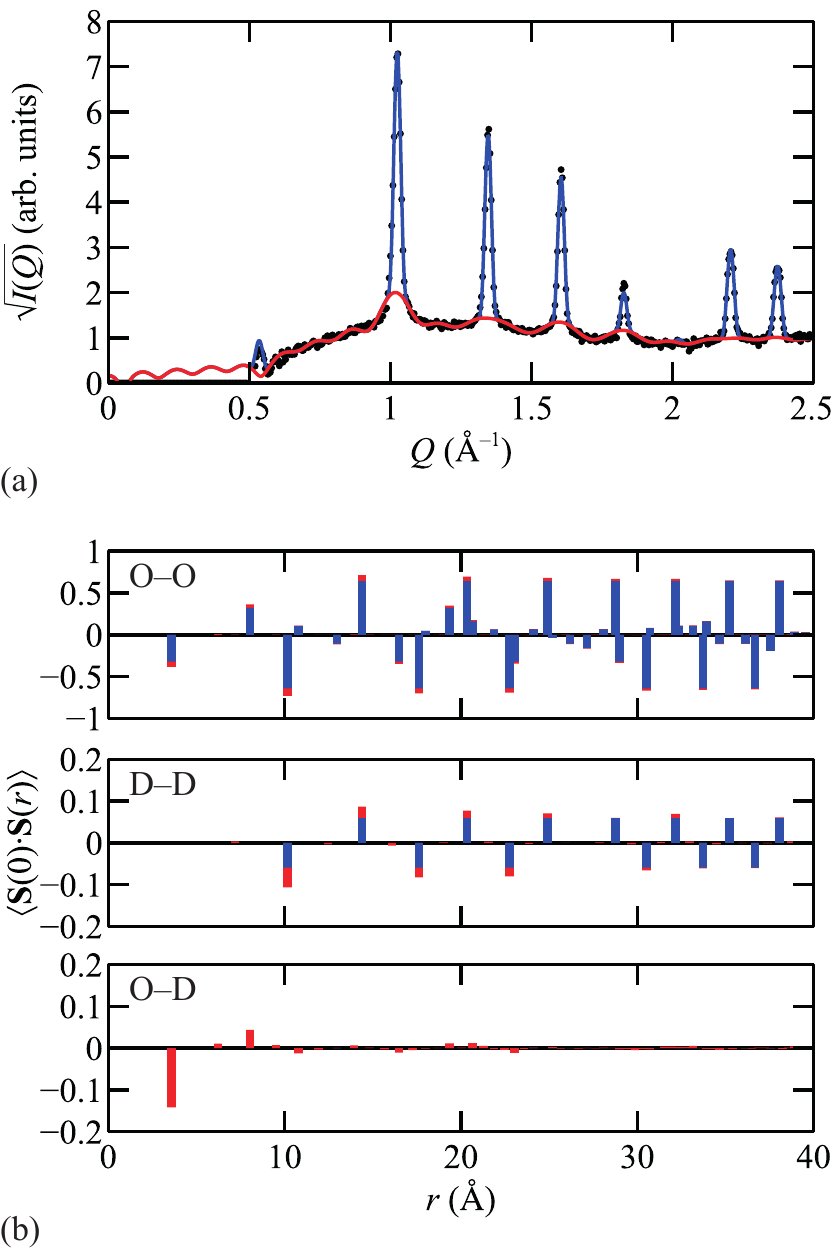}
\end{center}
\caption{\label{rmc_fits} (a) Reverse Monte Carlo (RMC) fit to experimental magnetic total-scattering data from Ref.~\onlinecite{Stewart_2004}. Experimental data are shown as black circles, the Bragg component of the RMC fit as a blue line, and the diffuse component of the RMC fit as a red line. A square-root scale is used on the vertical axis to highlight the weak diffuse scattering. (b) Partial spin correlation functions. Panels show correlations between ordered--ordered (``O--O"), disordered--disordered (``D--D"), and ordered--disordered (``O--D") sites. In each panel, average-structure correlations are shown as blue bars and spin-fluctuation correlations are shown as red bars. Note the different vertical scales in each panel. The cubic lattice parameter $a=10.17$\,\AA.}\end{figure}

The RMC fit is shown in Fig.~\ref{rmc_fits}(a). The excellent quality of the fit proves that the magnetic diffuse scattering is consistent with the 1-$\mathbf{k}$ structure. Refinements were also performed where the average structure was constrained to be 4-$\mathbf{k}$, which yielded similarly good fits (not shown). The diffuse scattering shows two distinct components. First, there is a broad feature centred at $Q \approx 1.1\,$\AA$^{-1}$ which arises from antiferromagnetic spin correlations at $r \approx 3.6$\,\AA~\cite{Stewart_2004}; second, sharper features are also observed at the base of magnetic Bragg peaks. To understand the origin of these features we calculate partial spin correlation functions,
\begin{equation}
\left\langle \mathbf{S}(0)\cdot\mathbf{S}(r)\right\rangle _{\rho\sigma}=\frac{1}{N_{\rho}Z_{\rho\sigma}(r)}\sum_{\rho}^{N_{\rho}}\sum_{\sigma}^{Z_{\rho\sigma}(r)}\mathbf{S}_{i}^{\rho}\cdot\mathbf{S}_{j}^{\sigma},
\end{equation}
where $\rho,\sigma$ label either ordered or disordered spins, $N_\rho$ is the number of spins of type $\rho$, $Z_{\rho\sigma}(r)$ is the number of spins of type $\sigma$ which coordinate a spin of type $\rho$ at distance $r$, and spins $\mathbf{S}_i$ are normalised to unit length. We split the calculated $\left\langle \mathbf{S}(0)\cdot\mathbf{S}(r)\right\rangle _{\rho\sigma}$ into two parts: the average-structure correlation function, which does not decay with distance and gives rise to Bragg scattering, and the spin-fluctuation correlation function, which decays with distance and gives rise to diffuse scattering. These quantities are shown in Fig.~\ref{rmc_fits}(b). The magnitude of average-structure correlations is much greater for the ordered than the disordered site, as anticipated from the relative values of $\mu_{\mathrm{ord}}$ \cite{Stewart_2004}. The spin fluctuations resemble the average structure and persist over several crystallographic unit cells, consistent with a superposition of low-energy spin waves \cite{Mellergard_1999}. 

While the intra-site spin correlations account for the relatively sharp diffuse scattering at the base of the magnetic Bragg peaks, they are too long-ranged to account for the broad diffuse peak at $Q \approx 1.1$\,\AA$^{-1}$. Strikingly, however, the inter-site correlation function shows a significant antiferromagnetic peak at $r=3.6$\,\AA. We therefore propose that the broad feature arises from correlated spin fluctuations \emph{between} the ordered and disordered sites of the 1-$\mathbf{k}$ structure, rather than correlations within the disordered site of the 4-$\mathbf{k}$ structure as was previously proposed \cite{Stewart_2004}. Importantly, this result shows that the ordered and disordered sites are not independent, as previous studies suggested \cite{Champion_2001,Stewart_2004}, but are actually strongly coupled. However, this coupling can only occur via spin fluctuations, because symmetry constrains the inter-site correlations of the average structure to be zero. While our neutron-diffraction data do not directly determine the timescale of these fluctuations, neutron spin echo \cite{Ehlers_2006} and muon-spin rotation  \cite{Yaouanc_2005,Dunsiger_2006} measurements have shown that low-energy spin dynamics persist at $T \ll T_2$, suggesting that they are dynamic rather than frozen.

Our determination of the nature of partial order in Gd$_{2}$Ti$_{2}$O$_{7}$ provides an experimental foundation for theoretical studies which aim to understand the microscopic origins of partial-order states \cite{Javanparast_2015,Chern_2008}. Our key result is that the magnetic structure of Gd$_{2}$Ti$_{2}$O$_{7}$ for $T\ll T_2$ is not 4-$\mathbf{k}$, but the 1-$\mathbf{k}$ structure is consistent with all our data. We note that our results apply only for $T\ll T_2$, and it remains possible that a different magnetic structure could exist for $T_{2}<T<T_{1}$, as was recently proposed in Ref.~\onlinecite{Javanparast_2015}.  We anticipate that the coupling of disordered and ordered sites via spin fluctuations may prove important in directing the nature of the magnetic ground state. Inelastic neutron-scattering experiments could provide more detailed information on the nature of this coupling, but are extremely challenging due to the high neutron-absorption cross-section of isotopically-natural Gd. 

\section{Acknowledgements}

J.A.M.P., A.B.C., and A.L.G. gratefully acknowledge financial support from the STFC, EPSRC (EP/G004528/2) and ERC (Ref: 279705). G.E. acknowledges funding by the Scientific User Facilities Division, Office of Basic Energy Sciences, U.S. Department of Energy. We are grateful to S.~T.~Bramwell, A.~S.~Wills, M.~J.~P.~Gingras, M.~J.~Cliffe, and P.~J.~Saines for useful discussions, to O.~Kirichek and the ISIS Sample Environment Group for cryogenic support, and to J.~Makepeace for assistance with \textsc{Topas} 5. Experiments at the ISIS Neutron and Muon Source were supported by a beam time allocation from the STFC (U.K.). 

\bibliography{jamp_full_refs}
\bibliographystyle{Science}

\end{document}